\def\aap{\ {A\&A}\ }

\def\aj{\ {AJ}\ }

\def\apj{\ {ApJ}\ }
\def\apjl{\ {ApJL}\ }
\def\apjs{\ {ApJS}\ }

\def\mnras{\ {MNRAS}\ }
\def\nat{\ {Nat}\ }

\def\pasj{\ {Publ. Astr. Soc. Japan}\ }

\documentclass[]{elsart}
\usepackage{natbib}
\usepackage{epsfig}

\usepackage{graphicx}

  \newcommand{\msun}{\mbox{${\rm M}_\odot$}}

\def\apgt{\ {\raise-.5ex\hbox{$\buildrel>\over\sim$}}\ } 
\def\aplt{\ {\raise-.5ex\hbox{$\buildrel<\over\sim$}}\ } 
\def\lt{\ {\raise-.5ex\hbox{$\buildrel>$}}\ } 
\def\gt{\ {\raise-.5ex\hbox{$\buildrel<$}}\ }

\begin{document}

\begin{frontmatter}

\title{A multiphysics and multiscale software environment for 
       modeling astrophysical systems}

\author[UvA]{Simon Portegies Zwart}
\author[Drexel]{Steve McMillan}
\author[UvA]{Stefan Harfst}
\author[UvA]{Derek Groen}
\author[Tokyo]{Michiko Fujii}
\author[UvA]{Breannd\'an \'O Nuall\'ain}
\author[UU]{Evert Glebbeek}
\author[Eddinburgh]{Douglas Heggie}
\author[Allegheny]{James Lombardi}
\author[IAS]{Piet Hut}
\author[UvA]{Vangelis Angelou}
\author[Tata]{Sambaran Banerjee}
\author[VUB]{Houria Belkus}
\author[NWU]{Tassos Fragos}
\author[NWU]{John Fregeau}
\author[UvA]{Evghenii Gaburov}
\author[UU]{Rob Izzard}
\author[IAS]{Mario Juri\'c}
\author[Oxford]{Stephen Justham}
\author[UvA]{Andrea Sottoriva}
\author[Maryland]{Peter Teuben}
\author[SaintMary]{Joris van Bever}
\author[TelAviv]{Ofer Yaron}
\author[UCSC]{Marcel Zemp}

\address[UvA]       {University of Amsterdam, Amsterdam, The Netherlands}
\address[Drexel]    {Drexel University  Philadelphia, PA, USA}
\address[Tokyo]     {University of Tokyo, Tokyo, Japan}
\address[UU]        {Utrecht University  Utrecht, the Netherlands}
\address[Eddinburgh]{University of Edinburgh  Edinburgh, UK}
\address[Allegheny] {Allegheny College   Meadville, PA, USA}
\address[IAS]       {Institute for Advanced Study  Princeton, USA}
\address[Tata]      {Tata Institute of Fundamental Research   India}
\address[VUB]       {Vrije Universiteit Brussel  Brussel, Belgium}
\address[NWU]       {Northwestern University  Evanston, IL, USA}
\address[Oxford]    {University of Oxford  Oxford, UK}
\address[Maryland]  {University of Maryland   College Park, MD, USA}
\address[SaintMary] {Saint Mary's University  Halifax, Canada}
\address[TelAviv]   {Tel Aviv University  Tel Aviv, Israel}
\address[UCSC]      {University of California Santa Cruz  Santa Cruz, CA, USA}

\date{Accepted 2008 ???. Received 2008 ???; in original form 2008 ???}
\pubyear{1687}
\maketitle
\label{firstpage}

\maketitle

\begin{abstract}

We present MUSE, a software framework for combining existing
computational tools for different astrophysical domains into a single
multiphysics, multiscale application. MUSE facilitates the coupling of
existing codes written in different languages by providing
inter-language tools and by specifying an interface between each
module and the framework that represents a balance between generality
and computational efficiency.  This approach allows scientists to use
combinations of codes to solve highly-coupled problems without the
need to write new codes for other domains or significantly alter their
existing codes.  MUSE currently incorporates the domains of stellar
dynamics, stellar evolution and stellar hydrodynamics for studying
generalized stellar systems.  We have now reached a ``Noah's Ark''
milestone, with (at least) two available numerical solvers for each
domain.  MUSE can treat multi-scale and multi-physics systems in which
the time- and size-scales are well separated, like simulating the
evolution of planetary systems, small stellar associations, dense
stellar clusters, galaxies and galactic nuclei.
In this paper we describe three examples calculated using MUSE: the
merger of two galaxies, the merger of two evolving stars, and a hybrid
$N$-body simulation.  In addition, we demonstrate an implementation of
MUSE on a distributed computer which may also include special-purpose
hardware, such as GRAPEs or GPUs, to accelerate computations.  The
current MUSE code base is publicly available as open source at {\tt
http://muse.li}.

\end{abstract}
\end{frontmatter}

\section{Introduction}
 
The Universe is a multi-physics environment in which, from an
astrophysical point of view, Newton's gravitational force law,
radiative processes, nuclear reactions and hydrodynamics mutually
interact.  The astrophysical problems which are relevant to this study
generally are multi-scale, with spatial and temporal scales ranging
from $10^4$ m and $10^{-3}$ seconds on the small end to $10^{20}$m and
$10^{17}$s on the large end. The combined multi-physics, multi-scale
environment presents a tremendous theoretical challenge for modern
science.  While observational astronomy fills important gaps in our
knowledge by harvesting ever-wider spectral coverage with continuously
increasing resolution and sensitivity, our theoretical understanding
lags behind these exciting developments in instrumentation.

In many ways, computational astrophysics lies intermediate between
observations and theory.  Simulations generally cover a wider range of
physical phenomena than observations with individual telescopes,
whereas purely theoretical studies are often tailored to solving sets
of linearized differential equations. As soon as these equations show
emergent behavior in which the mutual coupling of non-linear processes
result in complex behavior, the computer provides an enormous resource
for addressing these problems.  Furthermore simulations can support
observational astronomy by mimicking observations and aiding their
interpretation by enabling parameter-space studies.  They can
elucidate the often complex consequences of even simple physical
theories, like the non-linear behavior of many-body gravitational
systems.  But in order to deepen our knowledge of the physics,
extensive computer simulations require large programming efforts and a
thorough fundamental understanding of many aspects of the underlying
theory.

From a management perspective, the design of a typical simulation
package differs from construction of a telescope in one very important
respect.  Whereas modern astronomical instrumentation is generally
built by teams of tens or hundreds of people, theoretical models are
usually one-person endeavors.  Pure theory lends itself well to this
relatively individualistic approach, but scientific computing is in a
less favorable position.  So long as the physical scope of a problem
remains relatively limited, the necessary software can be built and
maintained by a single numerically educated astronomer or scientific
programmer.  However, these programs are often ``single-author,
single-use'', and thus single-purpose: recycling of scientific
software within astronomy is still rare.

More complex computer models often entail non-linear couplings between
many distinct, and traditionally separate, physical domains.
Developing a simulation environment suitable for multi-physics
scientific research is not a simple task.  Problems which encompass
multiple temporal or spatial scales are often coded by small teams of
astronomers.  Many recent successful projects have been carried out in
this way; examples are GADGET \citep{2001NewA....6...79S}, and starlab
\citep{2001MNRAS.321..199P}.  In all these cases, a team of scientists
collaborated in writing a large-scale simulation environment.  The
resulting software has a broad user base, and has been applied to a
wide variety of problems.  These packages, however, each address one
quite specific task, and their use is limited to a rather narrow
physical domain. In addition, considerable expertise is needed to use
them and expanding these packages to allow the addition of a new
physical domain is hampered by early design choices.

In this paper we describe a software framework that targets
multi-scale, multi-physics problems in a hierarchical and internally
consistent implementation.  Its development is based on the philosophy
of ``open knowledge'' \footnote{See for example {\tt
    http://www.artcompsci.org/ok/}.}.  We call this environment MUSE,
for MUltiphysics Software Environment.

\section{The concept of MUSE}

\begin{figure}
\psfig{figure=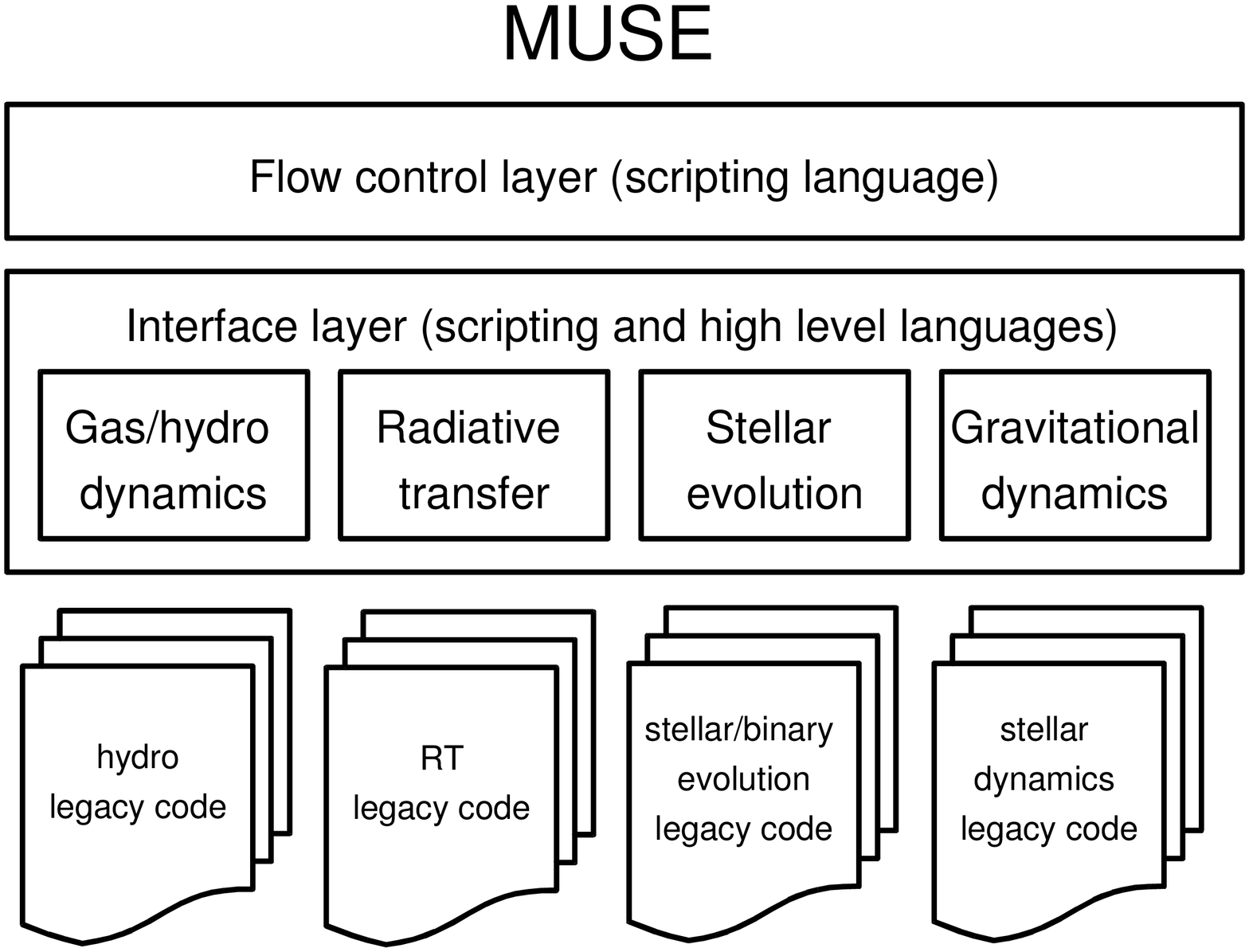,width=0.7\columnwidth,angle=-90}
\caption[]{ 
        Basic structure design of the framework (MUSE).  The top layer
        (flow control) is connected to the middle (interface layer)
        which controls the command structure for the individual
        applications. In this example only an indicative selection of
        numerical techniques is shown for each of the applications.
\label{Fig:amuse_structure}
}
\end{figure}

The development of MUSE began during the MODEST-6a\footnote{MODEST
stands for MOdeling DEnse STellar Systems; the term was coined during
the first MODEST meeting in New York (USA) in 2001.  The MODEST web
page is {\tt http://www.manybody.org/modest}; see also
\citet{2003NewA....8..337H,2003NewA....8..605S}.}  workshop in Lund,
Sweden \citep{2006NewA...12..201D}, but the first lines of code were
written during MODEST-6d/e in Amsterdam (the Netherlands).  The idea
of Noah's Ark (see \S\,\ref{Sect:Noah}) was conceived and realized in
2007, during the MODEST-7f workshop in Amsterdam and the MODEST-7a
meeting in Split (Croatia).

The development of a multi-physics simulation environment can be
approached from a monolithic or from a modular point of view.  In the
monolithic approach a single numerical solver is systematically
expanded to include more physics.  Basic design choices for the
initial numerical solver are generally petrified in the initial
architecture.  Nevertheless, such codes are sometimes successfully
redesigned to include two or possibly even three solvers for a
different physical phenomenon \citep[see FLASH, where hydrodynamics
has been combined with magnetic
fields][]{2000ApJS..131..273F}). Rather than forming a self-consistent
framework, the different physical domains in these environments are
made to co-exist. This approach is prone to errors, and the resulting
large simulation packages often suffer from bugs, redundancy in source
code, sections of dead code, and a lack of homogeneity.  The
assumptions needed to make these codes compile and operate without
fatal errors often hampers the science. In addition, the underlying
assumptions are rarely documented, and the resulting science is often
hard to interpret.

We address these issues in MUSE by the development of a modular
numerical environment, in which independently developed specialized
numerical solvers are coupled at a meta level, resulting in the
coherent framework as depicted in Fig.\,\ref{Fig:amuse_structure}.
Modules are designed with well defined interfaces governing their
interaction with the rest of the system.  Scheduling of, and
communication among modules is managed by a top-level ``glue''
language.  In the case of MUSE, this glue language is Python, chosen
for its rich feature set, ease of programming, object-oriented
capabilities, large user base, and extensive user-written software
libraries.  However, we have the feeling that Python is not always
consistent and of equally high quality in all places. The objective of
the glue code is to realize the interoperation between different parts
of the code, which may be realized via object-relational mapping, in
which individual modules are equipped with instruction sets to
exchange information with other modules.

The modular approach has many advantages.  Existing codes which have
been well tuned and tested within their own domains can be reused by
wrapping them in a thin interface layer and incorporating them into a
larger framework.  The identification and specification of suitable
interfaces for such codes allows them to be interchanged easily.  An
important element of this design is the provision of documentation and
exemplars for the design of new modules and for their integration into
the framework. A user can ``mix and match'' modules like building
blocks to find the most suitable combination for the application at
hand, or to compare them side by side.  The first interface standard
between stellar evolution and stellar dynamics goes back
to \cite{2003NewA....8..337H}.  The resulting software is also more
easily maintainable, since all dependencies between a module and the
rest of the system are well defined and documented.

A particular advantage of a modular framework is that a motivated
scholar can focus attention on a narrower area, write a module for it,
then integrate it into the framework with knowledge of only the bare
essentials of the interface.  For example, it would take little extra
work to adapt the results of a successful student project into a
separate module, or for a researcher working with a code in one field
of physics to find out how the code interacts in a multi-physics
environment.  The shallower learning curve of the framework
significantly lowers the barrier for entry, making it more accessible
and ultimately leading to a more open and extensible system.

The only constraint placed on a new module is that it must be written
(or wrapped) in a programming language with a Foreign Function
Interface that can be linked to a contemporary Unix-like system.  As
in the high-level language \texttt{Haskell}, a Foreign Function
Interface provide a mechanism by which a program written in one
language can call routines from another language. Supported languages
include low-level (\texttt{C},
\texttt{C++} and \texttt{Fortran}) as well as other high-level
languages such as \texttt{C\#}, \texttt{Java}, \texttt{Haskell},
\texttt{Python} and \texttt{Ruby}. Currently, individual MUSE modules
are written in \texttt{Fortran}, \texttt{C}, and \texttt{C++}, and are
interfaced with \texttt{Python} using \texttt{f2py} or \texttt{swig}.
Several interfaces are written almost entirely in Python, minimizing
the programming burden on the legacy programmer.  The flexibility of
the framework allows a much broader range of applications to be
prototyped, and the bottom-up approach makes the code much easier to
understand, expand and maintain.  If a particular combination of
modules is found to be particularly suited to an application, greater
efficiency can be achieved by hard coding the interfaces
and factoring out the glue code, thus effectively ramping up to a
specialized monolithic code.

\subsection{Noah's Ark}\label{Sect:Noah}

\begin{table*}
\caption[]{Modules currently present (or in preparation) in MUSE. The
  codes are identified by their acronym, which is also used on the
  MUSE repository at {\tt http://muse.li}, followed by a short
  description. Some of the modules mentioned here are used in
  \S\,\ref{Sect:Examples}.  Citations to the literature are indicated
  in the second column by their index 
  1:\citet{1989ApJ...347..998E}, 2:\citet{2006epbm.book.....E},
  3:\citet{1995ApJ...443L..93H},
  4:\citet{1992PASJ...44..141M}, 5:\citet{2007NewA...12..357H},
  6:\citet{1986Natur.324..446B}, 7:\citet{2003MNRAS.345..762L};
  8:\citet{2008HLA.RBS.I,2008HLA.RBS.II}; 
  9:\citet{2002ApJ...570..171F,2003ApJ...593..772F}; 
  10:\citet{2007PASJ...59.1095F}.
For a number of modules the source code is currently not available
within MUSE because they are not publicly available or still under
development. Those are the Henyey stellar evolution code
\texttt{EVTwin} \citep{1971MNRAS.151..351E,2006epbm.book.....E}, the
Monte-Carlo dynamics module \texttt{cmc}
\citep{2000ApJ...540..969J,2003ApJ...593..772F}, the hybrid $N$-body
integrator \texttt{BRIDGE} \citep[][used in
  \S\,\ref{Sect:Bridge}]{2007PASJ...59.1095F} and the Monte-Carlo
radiative transfer code MCRT.}
\bigskip
\begin{tabular}{llll}
\hline
MUSE module&ref.& language & brief description \\
\hline
EFT89         &1& C  & Parameterized stellar evolution \\       
EVTwin        &2& F77/F90& Henyey code to evolve stars \\   
Hermite0      &3& C++& Direct $N$-body integrator \\      
NBODY1h       &4& F77& Direct $N$-body integrator \\      
phiGRAPE      &5& F77& (parallel) direct $N$-body integrator \\      
BHTree        &6& C++& Barnes-Hut tree code \\   
SmallN        & & C++& Direct integrator for systems of few bodies \\   
Sticky Spheres& & C++& Angular momentum and energy conserving collision treatment\\
mmas          &7& F77& Entropy sorting for merging two stellar structures \\
MCRT          & & C++& Monte-Carlo Radiative Transfer \\
\hline
Globus support& & Python& Support for performing simulations on distributed resources \\
HLA grid support&8& HLA& Support for performing simulations on distributed resources \\
Scheduler     & & Python& Schedules the calling sequence between modules \\
Unit module   & & Python& Unit conversion \\
XML parser    & & Python& Primitive parser for XML formatted data \\
\hline
cmc           &9 & C  & Monte Carlo stellar dynamics module \\
BRIDGE        &10& C++& Hybrid direct $N$-body with Barnes-Hut Tree code \\
\hline
\label{Tab:MUSEmodules}
\end{tabular}
\end{table*}

Instead of writing all new code from scratch, in MUSE we realized a
software framework in which the glue language provides an
object-relational mapping via a virtual library which is used to bind
a wide collection of diverse applications.

MUSE consists of a hierarchical component architecture that
encapsulates dynamic shared libraries for simulating stellar
evolution, stellar dynamics and treatments for colliding stars.  As
part of the MUSE specification, each module manages its own internal
(application-specific) data, communicating through the interface only
the minimum information needed for it to interoperate with the rest of
the system.  Additional packages for file I/O, data analysis and
plotting are included. Our objective is eventually to incorporate
treatments of gas dynamics and radiative transfer, but at this point
these are not yet implemented.

We have so far included at least two working packages for each of the
domains of stellar collisions, stellar evolution and stellar dynamics,
in what we call the ``Noah's Ark'' milestone. The homogeneous interface
that connects the kernel modules enables us to switch packages at
runtime via the scheduler.  The goal of this paper is to demonstrate
the modularity and interchangeability of the MUSE framework.  In
Tab.\,\ref{Tab:MUSEmodules} we give an overview of the currently
available modules in MUSE.

\subsubsection{Stellar dynamics} 
To simulate gravitational dynamics (e.g. between stars and/or
planets), we incorporate packages to solve Newton's equations of
motion by means of gravitational $N$-body solvers.  Several distinct
classes of $N$-body kernels are currently available. These are based
on the direct force evaluation methods or tree codes.

Currently four direct $N$-body methods are incorporated, all of which
are based on the fourth-order Hermite predictor-corrector $N$-body
integrator, with block time steps \citep{1992PASJ...44..141M}. Some of
them can benefit from special-purpose hardware such as GRAPE
\citep{1998sssp.book.....M,2001ASPC..228...87M} or a GPU
\citep{2007NewA...12..641P,2008NewA...13..103B}.  Direct methods
provides the high accuracy needed for simulating dense stellar
systems, but even with special computer hardware they lack the
performance to simulate systems with more than $\sim10^6$ particles.
The Barnes-Hut tree-codes \citep{1986Natur.324..446B} are included for
use in simulations of large-$N$ systems. Two of the four codes are
also GRAPE/GPU-enabled.

\subsubsection{Stellar evolution} 
Many applications require the structure and evolution of stars to be
followed at various levels of detail.  At a minimum, the dynamical
modules need to know stellar masses and radii as functions of time,
since these quantities feed back into the dynamical evolution.  At
greater levels of realism, stellar temperatures and luminosities (for
basic comparison with observations), photon energy distributions (for
feedback on radiative transfer), mass loss rates, outflow velocities
and yields of various chemical elements (returned to the gas in the
system), and even the detailed interior structure (to follow the
outcome of a stellar merger or collision), are also important.
Consequently the available stellar evolution modules should ideally
include both a very rapid but approximate code for applications where
speed (enabling large numbers of stars) is paramount (e.g. when using
the Barnes-Hut tree code to follow the stellar dynamics) as well as a
detailed (but much slower) structure and evolution code where accuracy
is most important (for example when studying specific objects in
relatively small but dense star clusters).

Currently two stellar evolution modules are incorporated into
MUSE. One, called \texttt{EFT89}, is based on fits to pre-calculated
stellar evolution tracks \citep{1989ApJ...347..998E}, the other solves
the set of coupled partial differential equations that describe
stellar structure and evolution following the Henyey code originally
designed by \citet{1971MNRAS.151..351E}. The latter code, called
\texttt{EVTwin} is a fully implicit stellar evolution code that solves
the stellar structure equations and the reaction-diffusion equations
for the six major isotopes concurrently on an adaptive mesh
\citep{2008A&A...488.1007G}.  
EVTwin is designed to follow in detail the internal evolution of a
star of arbitrary mass. The basic code, written in Fortran 77/90,
operates on a single star---that is, the internal data structures
(Fortran common blocks) describe just one evolving object.
The \texttt{EVTwin} variant describes a pair of stars, the components
of a binary, and includes the possibility of mass transfer between
them.  A single star is modeled as a primary with a distant,
non-evolving secondary.  The lower speed of
\texttt{EVTwin} is inconvenient, but the much more flexible
description of the physics allows for more realistic treatment of
unconventional stars, such as collision products.

Only two \texttt{EVTwin} functions---the ``initialize'' and ``evolve''
operations---are exposed to the MUSE environment.  The F90 wrapper
also is minimal, providing only data storage and the commands needed
to swap stellar models in and out of \texttt{EVTwin} and to return
specific pieces of information about the stored data.  All other
high-level control structures, including identification of stars and
scheduling their evolution, is performed in a python layer that forms
the outer shell of the \texttt{EVTwin} interface. The result is that
the structure and logic of the original code is largely preserved,
along with the expertise of its authors.

\subsubsection{Stellar collisions} 

Physical interactions between stars are currently incorporated into
MUSE by means of one of two simplified hydrodynamic solvers.  The
simpler of the two is based on the ``sticky sphere'' approximation, in
which two objects merge when their separation becomes less than the
sum of their effective radii.  The connection between effective and
actual radius is calibrated using more realistic SPH simulations of
stellar collisions.  The second is based on the {\tt make-me-a-star}
(MMAS) package\footnote{See {\tt
http://webpub.allegheny.edu/employee/j/jalombar/mmas/}}
\citep{2003MNRAS.345..762L} and its extension {\tt
make-me-a-massive-star}\footnote{See {\tt
http://modesta.science.uva.nl/}} (MMAMS, \citet{2008MNRAS.383L...5G}).
MMA(M)S constructs a merged stellar model by sorting the fluid
elements of the original stars by entropy or density, then recomputing
their equilibrium configuration, using mass loss and shock heating
data derived from SPH calculations.  Ultimately, we envisage inclusion
of a full SPH treatment of stellar collisions into the MUSE framework.

MMAS (and MMAMS) can be combined with full stellar evolution models,
as they process the internal stellar structure in a similar fashion to
the stellar evolution codes.  The sticky sphere approximation only
works with parameterized stellar evolution, as it does not require any
knowledge of the internal stellar structure.

\subsubsection{Radiative transfer}

At this moment one example module for performing rudimentary radiative
transfer calculations is incorporated in MUSE. The module uses a
discrete grid of cells filled with gas or dust which is parameterized
in a local density $\rho$ and an opacity $\kappa$, with which we
calculate the optical depth ($\int \rho \kappa dx$). A star, that may
or may not be embedded in one of the grid cells emits $L$ photons,
each of which is traced through the medium until it is absorbed,
escapes or lands in the camera. In each cloud cell or partial cell a
photon has a finite probability that it is scattered or absorbed.
This probability is calculated by solving the scattering function $f$,
which depends on the angles and the Stokes parameter.  We adopt
electron scattering for gas and \cite{1941ApJ....93...70H} for dust
scattering (see \citep{2005MNRAS.362.1038E} for details).

\begin{figure}
%
\psfig{figure=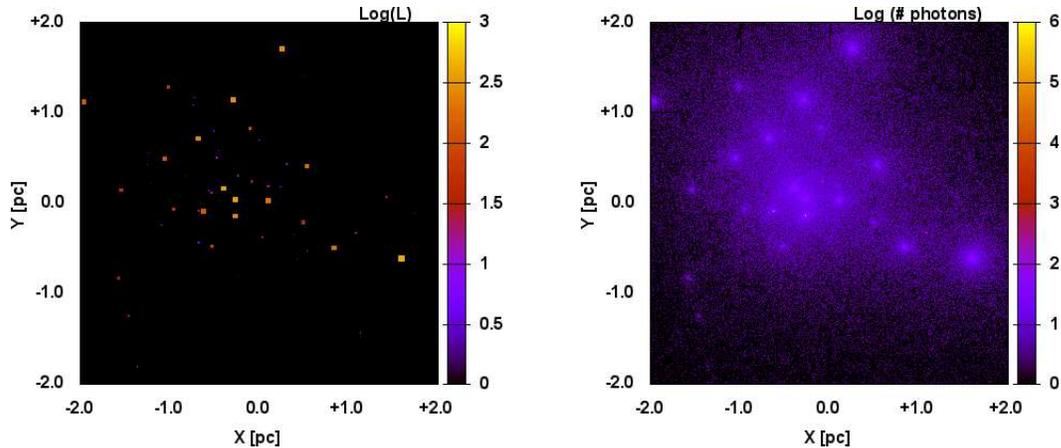,width=\columnwidth}
\caption[]{Radiative transfer module applied to a small $N=1024$
  particle Plummer sphere. Left image shows the intrinsic stellar
  luminosity at an age of 120\,Myr, the right image the image after
  applying the radiative transfer module for the cluster in a
  molecular cloud using a total of $10^7$ photon-packages.  The bar to
  the right of each frame indicates the logarithm of the luminosity of
  the star (left image) and the logarithm of the number of
  photons-packages that arrived in that particular pixel.
\label{Fig:RTPlummer}
}
\end{figure}

Since this module is in a rather experimental stage we only present
two images of its working, rather than a more complete description
in \S\,\ref{Sect:Examples}.  In Fig.\,\ref{Fig:RTPlummer} we present
the result of a cluster simulation using 1024 stars which initially
were distributed in a \cite{1911MNRAS..71..460P} sphere with a
virial radius of 1.32\,pc and in which the masses of the stars ware
selected randomly from the Salpeter
\citep{1955ApJ...121..161S} mass function between 1 and 100\,\msun,
totaling the cluster mass to about 750\,\msun.  These parameters ware
selected to mimic the Pleiades cluster \citep{2001MNRAS.321..199P}.
The cluster was scaled to virial equilibrium before we started its
evolution.  The cluster is evolved dynamically using the BHTree
package and the EFT89 module is used for evolving the stars.

We further assumed that the cluster was embedded in a giant molecular
cloud \citep{2000prpl.conf...97W}. The scattering parameters were set
to simulate visible light.  The gas and dust was distributed in a
homogeneous cube with 5pc on each side which was divided into
$1000\times1000\times100$ grid cells with a density of $10^2$ H$_2$
particles/cm$^3$. 

In Fig.\,\ref{Fig:RTPlummer} we present the central 5\,pc of the
cluster at an age of 120\,Myr.  The luminosity and position of the
stars are observed from the z-axis, i.e. they are projected on the
xy-plane. In the left panel we present the stellar luminosity color
coded, and the size of the symbols reflects the distance from the
observer, i.e., there it gives an indication of how much gas is
between the star and the observer. The right image is generated using
the MCRT module in MUSE and shows the photon-packages which were
traced from the individual stars to the camera position. Each
photon-package represents a multitude of photons.

\subsection{Units}

A notorious pitfall in combining scientific software is the failure to
perform correct conversion of physical units between modules.  In a
highly modular environment such as MUSE, this is a significant
concern.  One approach to the problem could have been to insist on a
standard set of units for modules incorporated into MUSE but
this is neither practical nor in keeping with the MUSE philosophy.

Instead, in the near future, we will provide a Units module in which
information about the specific choice of units the conversion factors
between them and certain useful physical constants are collected.
When a module is added to MUSE, the programmer adds a declaration of
the units it uses and expects.  When several modules are imported into
a MUSE experiment, the Units module then ensures that all external
values passed to each module are properly converted.

Naturally, the flexibility afforded by this approach also
introduces some overhead.  However, this very flexibility is MUSE's
great advantage; it allows an experimenter to easily mix and match
modules until the desired combination is found.  At that point, the
dependence on the Units module can be removed (if desired), and
conversion of physical units performed by explicit code.  This leads
to more efficient interfaces between modules, while the correctness of
the manual conversion can be checked at any time against the Units
module.

\subsection{Performance}

Large scale simulations, and in particular the multiscale and
multiphysics simulations for which our framework is intended, require
a large number of very different algorithms, many of which achieve
their highest performance on a specific computer architecture.  For
example, the gravitational $N$-body simulations are best performed on
a GRAPE enabled PC, the hydrodynamical simulations may be efficiently
accelerated using GPU hardware, while the trivially parallel
simultaneous modeling of a thousand single stars is best done on a
Beowulf cluster or grid computer.

The top-level organization of where each module should run is managed
using a resource broker, which is grid enabled (see
\S\,\ref{Sect:GRID}). We include a provision for individual packages to
indicate the class of hardware on which they operate optimally.  Some
modules are individually parallelized using the MPI library, whereas
others (like stellar evolution) are handled in a master-slave fashion
by the top-level manager.

\subsection{MUSE on the grid}\label{Sect:GRID}

Due to the wide range in computational characteristics of the available
modules, we generally expect to run MUSE on a computational grid
containing a number of specialized machines.  In this way we reduce
the run time by adopting computers which are best suited to each
module.  For example, we might select a large GRAPE cluster in Tokyo
for a direct $N$-body calculation, while the stellar evolution is
calculated on a Beowulf cluster in Amsterdam.  Here we report on our
preliminary grid interface, which allows us to use remote machines to
distribute individual MUSE modules.

The current interface uses the MUSE scheduler as the manager of grid
jobs and replaces internal module calls with a job execution
sequence. This is implemented with PyGlobus, an application
programming interface to the Globus grid middleware written in
Python. The execution sequence for each module consists of:
\begin{itemize}
 \item[$\bullet$] Write the state of a module, such as its initial conditions,
  to a file,
 \item[$\bullet$] transfer the state file to the destination site
 \item[$\bullet$] construct a grid job definition using the Globus
                  resource specification language
 \item[$\bullet$] submit the job to the grid; the grid job subsequently
  \begin{itemize}
   \item[-] reads the state file,
   \item[-] executes the specified MUSE module, 
   \item[-] writes the new state of the module to a file, and
   \item[-] copies the state file back to the MUSE scheduler
  \end{itemize}
 \item[$\bullet$] then read the new state file and resume the simulation.
\end{itemize}

The grid interface has been tested successfully using
DAS-3\footnote{see {\tt http://www.cs.vu.nl/das3/}}, which is a
five-cluster wide-area (in the Netherlands) distributed system
designed by the Advanced School for Computing and Imaging (ASCI).  We
executed individual invocations of stellar dynamics, stellar
evolution, and stellar collisions on remote machines.

\section{MUSE examples}\label{Sect:Examples}

\subsection{Temporal decomposition of two $N$-body codes}\label{Sect:TDecomp}

\begin{figure}
\psfig{figure=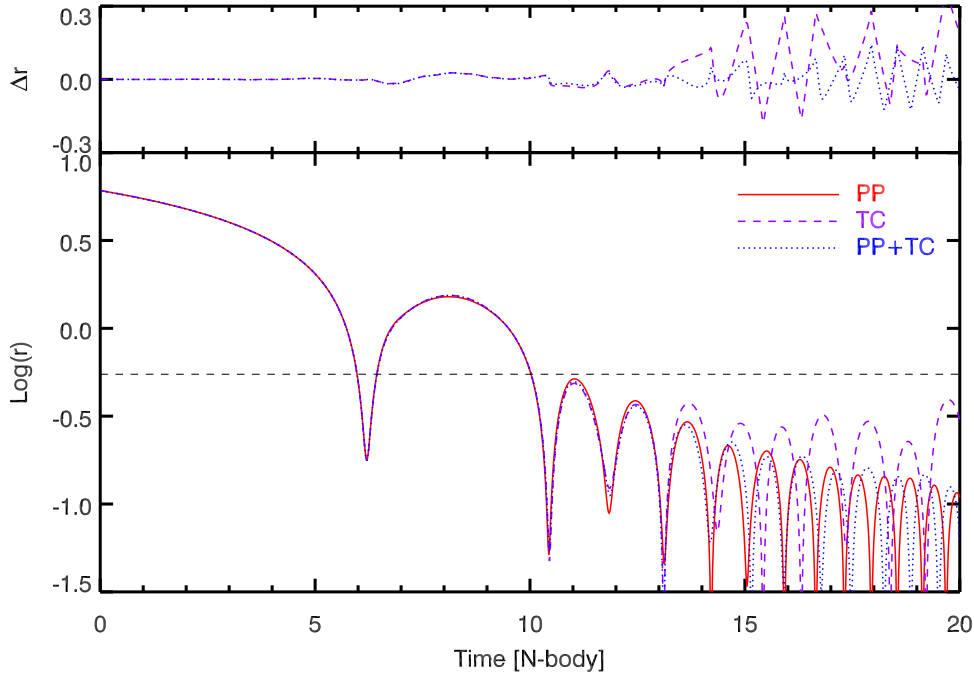,width=0.7\columnwidth,angle=90}
\caption[]{ 
  Time evolution of the distance between two black holes, each of
  which initially resides in the center of a ``galaxy,'' made up of
  32k particles, with total mass 100 times greater than the black hole
  mass.  Initially the two galaxies were far apart. The curves
  indicate calculations with the direct integrator (PP), a tree code
  (TC), and using the hybrid method in MUSE (PP+TC). The units along
  the axes are dimensionless $N$-body units
  \citep{1986LNP...267..233H}.
\label{Fig:gc_2k}
}
\end{figure}

Here we demonstrate the possibility of changing the integration method
within a MUSE application during runtime.  We deployed two integrators
to simulate the merging of two galaxies, each containing a central
black hole. The final stages of such a merger, with two black holes
orbiting one another, can only be integrated accurately using a direct
method.  Since this is computationally expensive, the early evolution
of such a merger is generally ignored and these calculations are
typically started some time during the merger process, for example
when the two black holes form a hard bound pair inside the merged
galaxy.

These rather arbitrary starting conditions for the binary black hole
merger problem can be improved by integrating the initial merger
between the two galaxies. We use the {\tt BHTree} code to reduce the
computational cost of simulating this merger process. At a
predetermined distance between the two black holes, when the tree code
is unlikely to produce accurate results, the simulation is continued
using the direct integration method for all particles. Overall this
results in a considerable reduction in runtime while still allowing an
accurate integration of the close black hole interaction.

Fig.\,\ref{Fig:gc_2k} shows the results of such a simulation.  The
initial conditions are two Plummer spheres, each consisting of 32k
equal-mass particles.  At the center of each ``galaxy'' we place a
black hole with mass 1\% that of the galaxy.  The two stellar systems
are subsequently set on a collision orbit, but at a fairly large
separation. We use two stellar dynamics modules (see
\S\ref{Sect:Noah}): \texttt{BHTree} \citep{1986Natur.324..446B}, with a
fixed shared time step, and \texttt{phiGRAPE}
\citep{2007NewA...12..357H}, a direct method using hierarchical block
time steps. Both modules are GRAPE-enabled and we make use of this to
speed up the simulation significantly. The calculation is performed
three times, once using {\tt phiGRAPE} alone, once using only {\tt
BHTree}, and once using the hybrid method. In the latter case the
equations of motion are integrated using {\tt phiGRAPE} if the two
black holes are within $\sim0.55$ N-body units\footnote{see {\tt
http://en.wikipedia.org/wiki/Natural\_units\#N-body\_units}.}
(indicated by the horizontal dashed line in Fig.\,\ref{Fig:gc_2k});
otherwise we use the tree code. Fig.\,\ref{Fig:gc_2k} shows the time
evolution of the distance between the two black holes.

The integration with only the direct {\tt phiGRAPE} integrator took
about 250 minutes, while the tree code took about 110 minutes. As
expected, the relative error in the energy of the direct $N$-body
simulation ($<10^{-6}$) is orders of magnitude smaller than the error
in the tree code ($\sim 1$\%). The hybrid code took about 200 minutes
to finish the simulation, with an energy error a factor of $\sim10$
better than that of the tree code. If we compare the time evolution of
the black hole separation for the tree and the hybrid code we find
that the hybrid code reproduces the results of the direct integration
(assuming it to be the most ``correct'' solution) quite well. In
summary, the hybrid method seems to be well suited to this kind of
problem, as it produces more accurate results than the tree code alone
and it is also faster than the direct code. The gain in performance is
not very large (only $\sim20$\%) for the particular problem studied
here, but as the compute time for the tree code scales with $N\log N$
whereas the direct method scales with $N^2$; a better gain is to be
expected for larger $N$. In addition, the MUSE implementation of the
tree code is rather basic, and both its performance and its accuracy
could be improved by using a variable block time step.

\subsection{Stellar mergers in MUSE}\label{Sect:StellarMergers}

Hydrodynamic interactions such as collisions and mergers can strongly
affect the overall energy budget of a dense stellar cluster and even
alter the timing of important dynamical phases such as core
collapse. Furthermore, stellar collisions and close encounters are
believed to produce a number of non-canonical objects, including blue
stragglers, low-mass X-ray binaries, recycled pulsars, double neutron
star systems, cataclysmic variables and contact binaries. These stars
and systems are among the most challenging to model and are also among
the most interesting observational markers. Predicting their numbers,
distributions and other observable characteristics is essential for
detailed comparisons with observations.

\begin{figure}
\psfig{figure=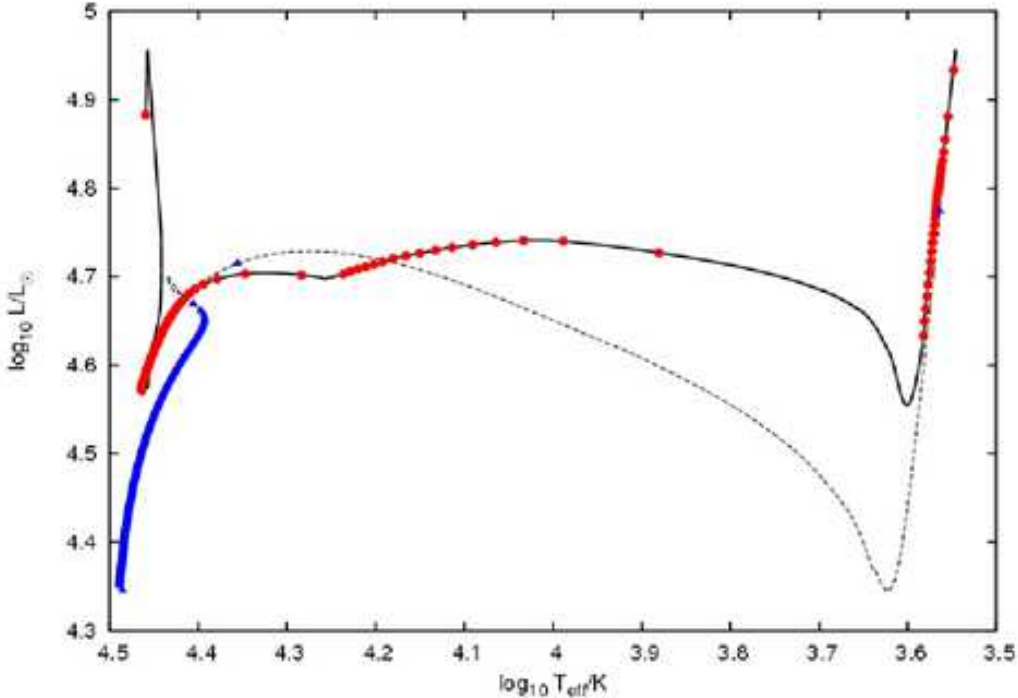,width=0.7\columnwidth,angle=270}
\caption{Evolution of a merger product formed by a collision between a
  $10 M_\odot$ star at the end of its main-sequence lifetime and a $7
  M_\odot$ star of the same age (filled circles), compared to the track of
  normal star of the same mass (15.7\,\msun) (triangles).  A symbol
  is plotted every $5\times 10^4$\,yr.  Both stars start their
  evolution at the left of the diagram (around $T_{\rm eff} \simeq
  3\times 10^4$\,K).  }\label{fig:hrd}
\end{figure}

When the stellar dynamics module in MUSE identifies a collision, the
stellar evolution module provides details regarding the evolutionary
state and structure of the two colliding stars. This information is
then passed on to the stellar collision module, which calculates the
structure of the merger remnant, returning it to the stellar evolution
module, which then continues its evolution.  This detailed treatment
of stellar mergers requires a stellar evolution module and a collision
treatment which resolve the internal structure of the stars; there is
no point in using a sticky-sphere approach in combination with a
Henyey-type stellar evolution code.

Fig.\,\ref{fig:hrd} presents the evolutionary track of a test case in
which \texttt{EVTwin} \citep{1971MNRAS.151..351E} (generally the more
flexible \texttt{TWIN} code is used, which allows the evolution of two
stars in a binary) was used to evolve the stars in our experiment.  A
10\,\msun\, star near the end of its main-sequence collided with a
7\,\msun\, star of the same age.  The collision itself was resolved
using \texttt{MMAMS}. The evolution of the resulting collision product
continued using \texttt{EVTwin}, which is presented as the solid curve
in Fig.\,\ref{fig:hrd}. For comparison we also plot (dashed curve) the
evolutionary track of a star with the same mass as the merger product.
The evolutionary tracks of the two stars are quite different, as are
the timescales on which the stars evolve after the main sequence and
through the giant branch.

The normal star becomes brighter as it follows an ordinary
main-sequence track, whereas the merged star fades dramatically as it
re-establishes thermal equilibrium shortly after the collision. The
initial evolution of the merger product is numerically difficult, as
the code attempts to find an equilibrium evolutionary track, which is
hard because the merger product has no hydrogen in its core. As a
consequence, the star leaves the main-sequence almost directly after
it establishes equilibrium, but since the core mass of the star is
unusually small (comparable to that of a 10\,\msun\, star at the
terminal-age main sequence) it is under luminous compared to the
normal star.  The slight kink in the evolutionary track between
$\log_{10} T_{\rm eff} = 4.2$ and 4.3 occurs when the merger product
starts to burn helium in its core. The star crosses the
Hertzsprung gap very slowly (in about 1 Myr), whereas the normal star
crosses within a few 10,000 years. This slow crossing is caused by the
small core of the merger product, which first has to grow to a mass to
be consistent with a $\sim 15.7$\,\msun\, star before it can leave
the Hertzsprung gap. The episode of extended Hertzsprung gap lifetime
is interesting as observing an extended lifetime Hertzsprung gap star
is much more likely than witnessing the actual collision.  Observing a
star on the Hertzsprung gap with a core too low in mass for its
evolutionary phase would therefore provide indirect evidence for the
collisional history of the star (regretfully one would probably
require some stellar stethoscope to observe the stellar core in such a
case).


\subsection{Hybrid $N$-body simulations with stellar evolution}\label{Sect:Bridge}

Dense star clusters move in the potential of a lower density
background. For globular clusters this is the parent's galaxy halo,
for open star clusters and dense young clusters it is the galactic
disc, and nuclear star clusters are embedded in their galaxy's bulge.
These high-density star clusters are preferably modeled using precise
and expensive direct-integration methods. For the relatively low
density regimes, however, a less precise method would suffice; saving
a substantial amount of compute time and allowing a much larger number
of particles to simulate the low-density host environment.  In
\S\,\ref{Sect:TDecomp} we described a temporal decomposition of a
problem using a tree code {\large O}($N\log(N)$) and a direct $N$-body
method.  Here we demonstrate a spatial domain decomposition using the
same methods.

The calculations performed in this \S\, are run via a MUSE module
which is based on BRIDGE \citep{2007PASJ...59.1095F}.  Within BRIDGE a
homogeneous and seamless transition between the different numerical
domains is possible with a similar method as is used in the
mixed-variable symplectic
method \citep{1991CeMDA..50...59K,1991AJ....102.1528W}, in which the
Hamiltonian is divided into two parts: an analytic Keplerian part and
the individual interactions between particles.  The latter are used to
perturb the regular orbits.  In our implementation the accurate direct
method, used to integrate the high-density regions, is coupled to the
much faster tree-code, which integrates the low-density part of the
galaxies. The stars in the high-density regions are perturbed by the
particles in the low-density environment.

The method implemented in MUSE and presented here uses an accurate
direct $N$-body solver (like {\tt Hermite0}) for the high density
regime whereas the rest of the system is integrated using {\tt
  BHTree}. In principle, the user is free to choose the integrator
used in the accurate part of the integration; in our current
implementation we adopt {\tt Hermite0}, but the tree-code is currently
petrified in the scheduler. This version of BRIDGE is currently not
available in the public version of MUSE.

To demonstrate the working of this hybrid scheme we simulate the
evolution of a star cluster orbiting within a galaxy.  The star
cluster is represented by 8192 particles with a Salpeter
\citep{1955ApJ...121..161S} mass function between 1 and 100 $M_{\odot}$
distributed according to a $W_0=10$\,King
model \citep{1966AJ.....71...64K} density profile.  This cluster is
located at a distance of 16\,pc from the center of the galaxy, with a
velocity of 65 km s$^{-1}$ in the transverse direction.  The galaxy is
represented by $10^6$ equal-mass particles in a $W_0=3$ King model
density distribution.  The stars in the star cluster are evolved using
the MUSE stellar evolution module \texttt{EFT89}, the galaxy particles
have all the same mass of 30\,\msun\ and were not affected by stellar
evolution.

The cluster, as it evolves internally, spirals in towards the galactic
center due to dynamical friction. While the cluster spirals in, it
experiences core collapse. During this phase many stars are packed
together in the dense cluster core and stars start to collide with
each other in a collision runaway process
\citep{1999A&A...348..117P}. These collisions are handled internally
in the direct part of BRIDGE. Throughout the core collapse of the
cluster about a dozen collisions occur with the same star, causing it
to grow in mass to about 700\,\msun. Although the stellar evolution of
such collision products is highly uncertain
\citep{2007ApJ...659.1576B,2007ApJ...668..435S}, we assume here that
the massive star collapses to a black hole of intermediate mass.

The direct part as well as the tree-part of the simulation was
performed on a full 1~Tflops GRAPE-6 \citep{2003PASJ...55.1163M},
whereas the tree-code was run on the host PC. The total CPU time for
this simulation was about half a day, whereas without BRIDGE the run
would have taken years to complete. The majority ($\sim 90\%$) of the
compute time was spent in the tree code, integrating the $10^6$
particles in the simulated galaxy.  (Note again that this fraction
depends on the adopted models and the use of special-purpose hardware
to accelerate the direct part of the integration.) Total energy was
conserved to better than $2\times 10^{-4}$ (initial total energy was
-0.25).

\begin{figure}
 \begin{center}
\psfig{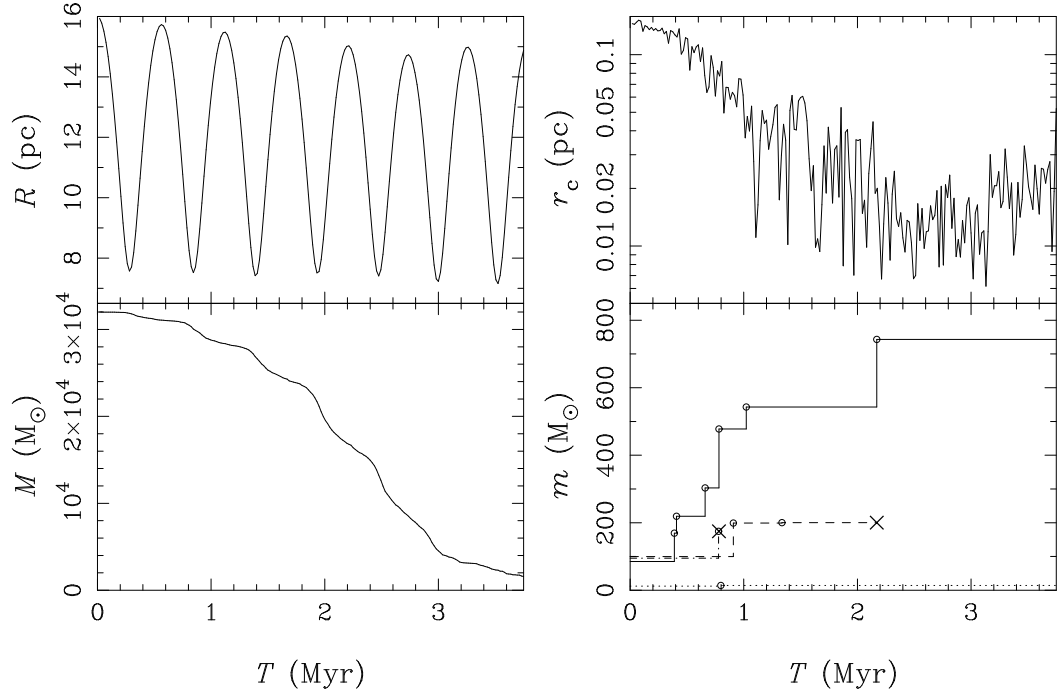}

 \caption{Results of the hybrid $N$-body simulation using a 4th-order
   Hermite scheme for the particles integrated directly and a
   Barnes-Hut tree algorithm for the others.
   The top left panel: distance from the cluster to the Galactic
   center,
   top right: evolution of the cluster core radius,
   bottom left: bound cluster mass, 
   bottom right: evolution of the mass of a few cluster stars that
   happen to experience collisions.
   The two crosses in the bottom right panel indicate the moment that
   two collision products coalesce with the runaway merger.  }
 \label{Bridge}
 \end{center}
\end{figure}

The results of the simulation are presented in
Fig.\,\ref{Bridge}. Here we see how the cluster (slightly) spirals in,
due to dynamical friction with the surrounding (tree-code) stars,
toward the galactic center before dissolving at an age of about
4\,Myr.  By that time, however, the runaway collision has already
resulted in a single massive star of more than 700\,\msun.

The description of stellar evolution adopted in this calculation is
rather simple and does not incorporate realistic mass loss, and it is
expected that the collision runaway will have a mass of $\sim
50$\msun\, by the time it collapses to a black hole in a supernova
explosion. The supernova itself may be unusually bright (possibly like
SN2006gy \citep{2007Natur.450..388P}) and may collapse to a relatively
massive black hole \citep{2004Natur.428..724P}. Similar collision
runaway results were obtained using direct $N$-body simulations using
{\tt starlab} \citep{2002ApJ...576..899P} and {\tt
NBODY} \citep{2004ApJ...613.1143B}, and with Monte-Carlo
\citep{2004ApJ...604..632G,2006MNRAS.368..141F} stellar dynamics simulations.

\subsection{Direct $N$-body dynamics with live stellar evolution}

While MUSE contains many self-contained dynamical modules, all of the
stellar evolution modules described thus far have relied on simple
analytical formulations or lookup formulae.  Here we present a new
simulation combining a dynamical integrator with a ``live'' stellar
evolution code, following the detailed internal evolution of stars in
real time as the dynamics unfolds.  A similar approach has been
undertaken by Ross Church, in his PhD thesis.  The novel ingredient in
this calculation is a MUSE interface onto the
\texttt{EVTwin} stellar evolution program \citep{2006epbm.book.....E},
modified for use within MUSE (see \S\,\ref{Sect:StellarMergers} for a
description).

In keeping with the philosophy of not rewriting existing working code,
in incorporating \texttt{EVTwin} into \texttt{MUSE}, we have made
minimal modifications to the program's internal structure.  Instead,
we wrap the program in a F90 data-management layer which maintains a
copy of the stellar data for each star in the system.  Advancing a
system of stars simply entails copying the chosen star into the
program and telling it to take a step.  EVTwin proceeds with the task
at hand, blissfully unaware that it is advancing different stellar
models at every invocation (see \S\,\ref{Sect:StellarMergers}).

\begin{figure}
\begin{centering}
\psfig{figure=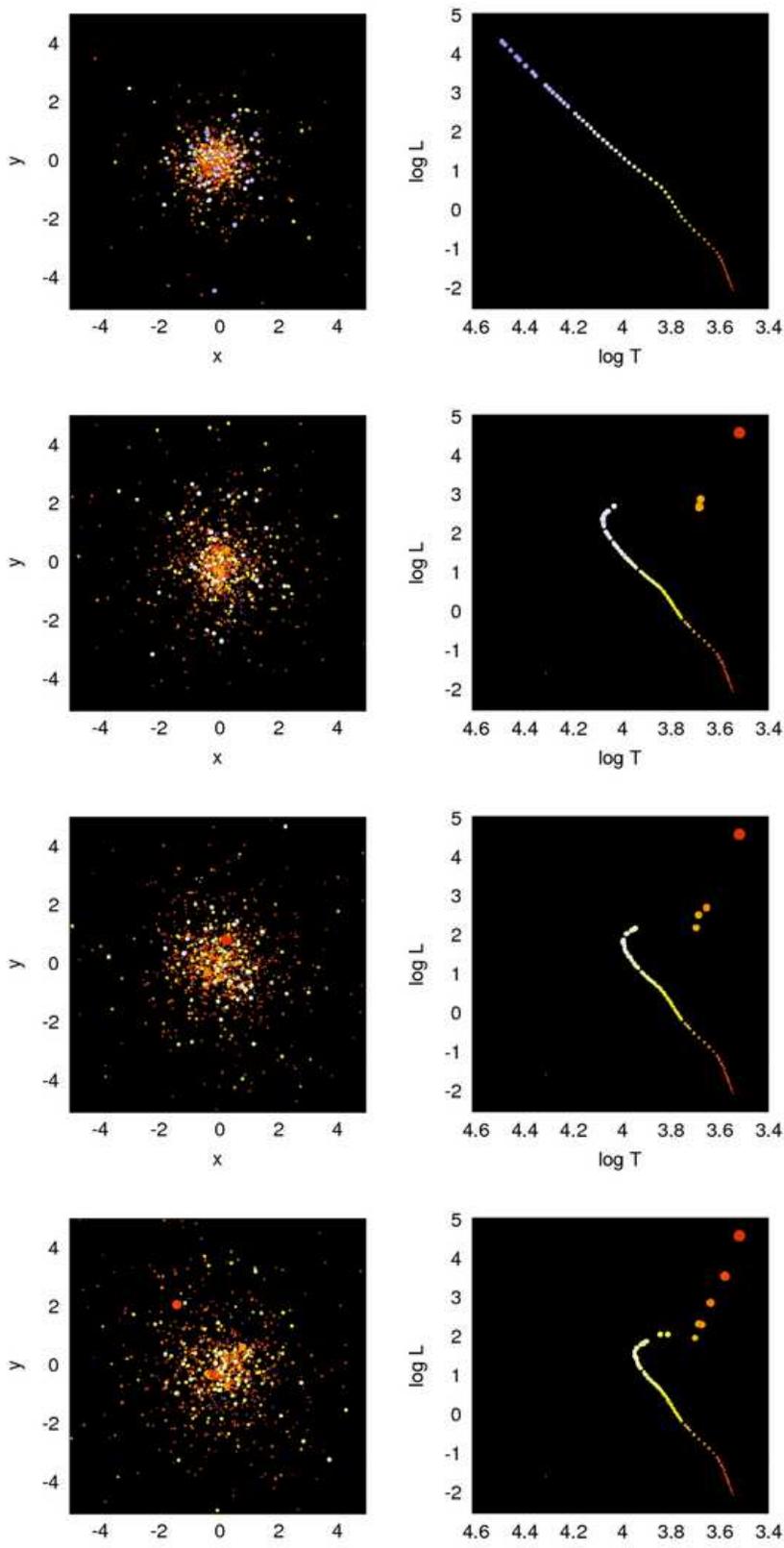,width=0.8\columnwidth,angle=0}
\vskip -0.25in
\end{centering}
\caption[]{Evolution of a 1024-body cluster, computed using the
{\tt hermite0} and {\tt EVTwin} MUSE modules.  The four rows of images
show the physical state of the cluster (left) and the cluster H--R
diagram (right) at times (top to bottom) 0, 200, 400, and 600 Myr.
Colors reflect stellar temperature, and radii are scaled by the
logarithm of the stellar radius.}
\label{Fig:stargrav}
\end{figure}

Figure \ref{Fig:stargrav} shows four snapshots during the evolution of
a 1024-body system, carried out within MUSE using EVTwin and the
simple shared-timestep {\tt hermite0} dynamical module.  Initially the
stars had a mass function $dN/dm \propto m^{-2.2}$ for $0.25 M_\odot <
m < 15 M_\odot$, for a mean mass of $0.92 M_\odot$ and were
distributed according to a Plummer density profile with a dynamical
time scale of 10 Myr, a value chosen mainly to illustrate the
interplay between dynamics and stellar evolution.  (The initial
cluster half-mass radius was $\sim 15$ pc.)  The initial half-mass
relaxation time of the system was 377 Myr.  The four frames show the
state of the system at times 0, 200, 400, and 600 Myr, illustrating
the early mass segregation and subsequent expansion of the system as
stars evolve and lose mass.

The integrator was kept deliberately simple, using a softened
gravitational potential to avoid the need for special treatment of
close encounters, and there was no provision for stellar collisions
and mergers.  Both collisions and close encounters will be added to
the simulation, and described in a future paper.  We note that,
although the {\tt hermite0} module is the least efficient member of
the MUSE dynamical suite, nevertheless the CPU time taken by the
simulation was roughly equally divided between the dynamical and
stellar modules.  Even without hardware acceleration (by GRAPE or
GPU), a more efficient dynamical integrator (such as one of the
individual block time step schemes already installed in MUSE) would
run at least an order of magnitude faster, underscoring the need for
careful load balancing when combining modules in a hybrid environment.

\section{Discussion}

The Multiscale Software Environment presented in this paper provides a
diverse and flexible framework for numerical studies of stellar
systems.  Now that the Noah's Ark milestone has been reached, one can
ask what new challenges MUSE has to offer.  Many of the existing
modules have been adapted for grid use and, as demonstrated in
\S\,\ref{Sect:GRID}, MUSE can be used effectively to connect various
computers around the world.  However, there are currently a number of
limitations in its use, and in its range of applications, which will
be addressed in the future.  Most of the current application modules
remain unsuitable for large-scale scientific production
simulations. The stellar dynamics codes do not yet efficiently deal
with close binaries and multiples, although modules are under
development, and external potentials, though relatively easy to
implement, have not yet been incorporated.  Binary evolution is not
implemented, and the diagnostics available to study the output of the
various modules remain quite limited.

Many improvements can be made to the environment, and we expect to
include many new modules, some similar to existing ones, others
completely different in nature.  The current framework has no method
for simulating interstellar gas, although this would be an extremely
valuable addition to the framework, enabling study of gas-rich star
clusters, galaxy collisions, colliding-wind binary systems, etc.  In
addition, radiation transfer is currently not implemented, nor are
radiative feedback mechanisms between stars and gas.  Both would
greatly increase the applicability base of the framework.  However,
both are likely to challenge the interface paradigm on which MUSE is
based.

The current MUSE setup, in which the individual modules are largely
decoupled, has a number of attractive advantages over a model in which
we allow direct memory access. The downside is that MUSE in its
present form works efficiently only for systems in which the various
scales are well separated.  Communication between the various modules,
even of the same type, is currently all done via the top interface
layer. For small studies, this poses relatively little overhead, but
for more extensive calculations, or those in which more detailed data
must be shared, it is desirable to minimize this overhead. One way to
achieve this would be by allowing direct data access between
modules. However, for such cases, the unit conversion modules could
not be used, and consistency in the units between the modules cannot
be guaranteed. As a result, each module would be required to maintain
consistent units throughout, which may be hard to maintain and prone
to bugs. In addition, the general problem of sharing data structures
between modules written in different languages, currently resolved by
the use of the glue language, resurfaces.

\section*{Acknowledgments}

We are grateful to Atakan G\"urkan, Junichiro Makino, Stephanie Rusli
and Dejan Vinkovi\'c for many discussions.  Our team meetings have
been supported by the Yukawa Institute for Theoretical Physics in
Kyoto, the International Space Science Institute in Bern, the
department of astronomy of the university of Split in Split, the
Institute for Advanced Study in Princeton and the Astronomical
Institute 'Anton Pannekoek' in Amsterdam.  This research was supported
in part by the Netherlands Organization for Scientific Research (NWO
grant No. 635.000.001 and 643.200.503), the Netherlands Advanced
School for Astronomy (NOVA), the Leids Kerkhoven-Bosscha fonds (LKBF),
the ASTROSIM program of the European Science Foundation, by NASA ATP
grants NNG04GL50G and NNX07AH15G, by the National Science Foundation
under grants AST-0708299 (S.L.W.M.) and PHY-0703545 (J.C.L.), by the
Special Coordination Fund for Promoting Science and Technology
(GRAPE-DR project), the Japan Society for the Promotion of Science
(JSPS) for Young Scientists, Ministry of Education, Culture, Sports,
Science and Technology, Japan and DEISA.  Some of the calculations
were done on the LISA cluster and the DAS-3 wide-area computer in the
Netherlands.  We are also grateful to SARA computing and networking
services, Amsterdam for their support.


\begin{thebibliography}{}

\bibitem[\protect\astroncite{{Barnes} \& {Hut}}{1986}]{1986Natur.324..446B}
{Barnes}, J., {Hut}, P. 1986, \nat, 324, 446

\bibitem[\protect\astroncite{{Baumgardt} et~al.}{2004}]{2004ApJ...613.1143B}
{Baumgardt}, H., {Makino}, J., {Ebisuzaki}, T. 2004, \apj, 613, 1143

\bibitem[\protect\astroncite{{Belkus} et~al.}{2007}]{2007ApJ...659.1576B}
{Belkus}, H., {Van Bever}, J., {Vanbeveren}, D. 2007, \apj, 659, 1576

\bibitem[\protect\astroncite{{Belleman} et~al.}{2008}]{2008NewA...13..103B}
{Belleman}, R.~G., {B{\'e}dorf}, J., {Portegies Zwart}, S.~F. 2008, New
  Astronomy, 13, 103

\bibitem[\protect\astroncite{{Davies} et~al.}{2006}]{2006NewA...12..201D}
{Davies}, M.~B., {Amaro-Seoane}, P., {Bassa}, C., {Dale}, J., {de Angeli}, F.,
  {Freitag}, M., {Kroupa}, P., {Mackey}, D., {Miller}, M.~C., {Portegies
  Zwart}, S. 2006, New Astronomy, 12, 201

\bibitem[\protect\astroncite{{Eggleton}}{2006}]{2006epbm.book.....E}
{Eggleton}, P. 2006,
\newblock Evolutionary Processes in Binary and Multiple Stars, 
  ISBN 0521855578, Cambridge University Press.

\bibitem[\protect\astroncite{{Eggleton}}{1971}]{1971MNRAS.151..351E}
{Eggleton}, P.~P. 1971, \mnras, 151, 351

\bibitem[\protect\astroncite{{Eggleton} et~al.}{1989}]{1989ApJ...347..998E}
{Eggleton}, P.~P., {Fitchett}, M.~J., {Tout}, C.~A. 1989, \apj, 347, 998

\bibitem[\protect\astroncite{{Ercolano} et~al.}{2005}]{2005MNRAS.362.1038E}
{Ercolano}, B., {Barlow}, M.~J., {Storey}, P.~J. 2005, \mnras, 362, 1038

\bibitem[\protect\astroncite{{Fregeau} et~al.}{2003}]{2003ApJ...593..772F}
{Fregeau}, J.~M., {G{\" u}rkan}, M.~A., {Joshi}, K.~J., {Rasio}, F.~A. 2003,
  \apj, 593, 772

\bibitem[\protect\astroncite{{Fregeau} et~al.}{2002}]{2002ApJ...570..171F}
{Fregeau}, J.~M., {Joshi}, K.~J., {Portegies Zwart}, S.~F., {Rasio}, F.~A.
  2002, \apj, 570, 171

\bibitem[\protect\astroncite{{Freitag} et~al.}{2006}]{2006MNRAS.368..141F}
{Freitag}, M., {G{\"u}rkan}, M.~A., {Rasio}, F.~A. 2006, \mnras, 368, 141

\bibitem[\protect\astroncite{{Fryxell} et~al.}{2000}]{2000ApJS..131..273F}
{Fryxell}, B., {Olson}, K., {Ricker}, P., {Timmes}, F.~X., {Zingale}, M.,
  {Lamb}, D.~Q., {MacNeice}, P., {Rosner}, R., {Truran}, J.~W., {Tufo}, H.
  2000, \apjs, 131, 273

\bibitem[\protect\astroncite{{Fujii} et~al.}{2007}]{2007PASJ...59.1095F}
{Fujii}, M., {Iwasawa}, M., {Funato}, Y., {Makino}, J. 2007, \pasj, 59, 1095

\bibitem[\protect\astroncite{{Gaburov} et~al.}{2008}]{2008MNRAS.383L...5G}
{Gaburov}, E., {Lombardi}, J.~C., {Portegies Zwart}, S. 2008, \mnras, 383, L5

\bibitem[\protect\astroncite{{Glebbeek} et~al.}{2008}]{2008A&A...488.1007G}
{Glebbeek}, E., {Pols}, O.~R., {Hurley}, J.~R. 2008, \aap, 488, 1007

\bibitem[\protect\astroncite{{G{\"u}rkan} et~al.}{2004}]{2004ApJ...604..632G}
{G{\"u}rkan}, M.~A., {Freitag}, M., {Rasio}, F.~A. 2004, \apj, 604, 632

\bibitem[\protect\astroncite{{Harfst} et~al.}{2007}]{2007NewA...12..357H}
{Harfst}, S., {Gualandris}, A., {Merritt}, D., {Spurzem}, R., {Portegies
  Zwart}, S., {Berczik}, P. 2007, New Astronomy, 12, 357

\bibitem[\protect\astroncite{{Heggie} \& {Mathieu}}{1986}]{1986LNP...267..233H}
{Heggie}, D.~C., {Mathieu}, R.~D. 1986, LNP Vol.~267: The Use of Supercomputers
  in Stellar Dynamics, in P. Hut, S. McMillan (eds.), Lecture Not. Phys 267,
  Springer-Verlag, Berlin

\bibitem[\protect\astroncite{{Henyey} \&
  {Greenstein}}{1941}]{1941ApJ....93...70H}
{Henyey}, L.~G., {Greenstein}, J.~L. 1941, \apj, 93, 70

\bibitem[\protect\astroncite{{Hut} et~al.}{1995}]{1995ApJ...443L..93H}
{Hut}, P., {Makino}, J., {McMillan}, S. 1995, \apjl, 443, L93

\bibitem[\protect\astroncite{{Hut} et~al.}{2003}]{2003NewA....8..337H}
{Hut}, P., {Shara}, M.~M., {Aarseth}, S.~J., {Klessen}, R.~S., {Lombardi}, Jr.,
  J.~C., {Makino}, J., {McMillan}, S., {Pols}, O.~R., {Teuben}, P.~J.,
  {Webbink}, R.~F. 2003, New Astronomy, 8, 337

\bibitem[\protect\astroncite{{Joshi} et~al.}{2000}]{2000ApJ...540..969J}
{Joshi}, K.~J., {Rasio}, F.~A., {Portegies Zwart}, S. 2000, \apj, 540, 969

\bibitem[\protect\astroncite{{King}}{1966}]{1966AJ.....71...64K}
{King}, I.~R. 1966, \aj, 71, 64

\bibitem[\protect\astroncite{{Kinoshita} et~al.}{1991}]{1991CeMDA..50...59K}
{Kinoshita}, H., {Yoshida}, H., {Nakai}, H. 1991, Celestial Mechanics and
  Dynamical Astronomy, 50, 59

\bibitem[\protect\astroncite{{Lombardi} et~al.}{2003}]{2003MNRAS.345..762L}
{Lombardi}, J.~C., {Thrall}, A.~P., {Deneva}, J.~S., {Fleming}, S.~W.,
  {Grabowski}, P.~E. 2003, \mnras, 345, 762

\bibitem[\protect\astroncite{{Makino}}{2001}]{2001ASPC..228...87M}
{Makino}, J. 2001,
\newblock in S. {Deiters}, B. {Fuchs}, A. {Just}, R. {Spurzem}, R. {Wielen}
  (eds.), ASP Conf. Ser. 228: Dynamics of Star Clusters and the Milky Way,
  p.~87

\bibitem[\protect\astroncite{{Makino} \& {Aarseth}}{1992}]{1992PASJ...44..141M}
{Makino}, J., {Aarseth}, S.~J. 1992, \pasj, 44, 141

\bibitem[\protect\astroncite{{Makino} et~al.}{2003}]{2003PASJ...55.1163M}
{Makino}, J., {Fukushige}, T., {Koga}, M., {Namura}, K. 2003, \pasj, 55, 1163

\bibitem[\protect\astroncite{{Makino} \& {Taiji}}{1998}]{1998sssp.book.....M}
{Makino}, J., {Taiji}, M. 1998,
\newblock {Scientific simulations with special-purpose computers : The GRAPE
  systems},
\newblock Scientific simulations with special-purpose computers : The GRAPE
  systems /by Junichiro Makino \& Makoto Taiji.~Chichester ; Toronto : John
  Wiley \& Sons, c1998.

\bibitem[\protect\astroncite{{Plummer}}{1911}]{1911MNRAS..71..460P}
{Plummer}, H.~C. 1911, \mnras, 71, 460

\bibitem[\protect\astroncite{{Portegies Zwart}
  et~al.}{2004}]{2004Natur.428..724P}
{Portegies Zwart}, S.~F., {Baumgardt}, H., {Hut}, P., {Makino}, J., {McMillan},
  S.~L.~W. 2004, \nat, 428, 724

\bibitem[\protect\astroncite{{Portegies Zwart}
  et~al.}{2007}]{2007NewA...12..641P}
{Portegies Zwart}, S.~F., {Belleman}, R.~G., {Geldof}, P.~M. 2007, New
  Astronomy, 12, 641

\bibitem[\protect\astroncite{{Portegies Zwart}
  et~al.}{1999}]{1999A&A...348..117P}
{Portegies Zwart}, S.~F., {Makino}, J., {McMillan}, S.~L.~W., {Hut}, P. 1999,
  \aap, 348, 117

\bibitem[\protect\astroncite{{Portegies Zwart} \&
  {McMillan}}{2002}]{2002ApJ...576..899P}
{Portegies Zwart}, S.~F., {McMillan}, S.~L.~W. 2002, \apj, 576, 899

\bibitem[\protect\astroncite{{Portegies Zwart}
  et~al.}{2001}]{2001MNRAS.321..199P}
{Portegies Zwart}, S.~F., {McMillan}, S.~L.~W., {Hut}, P., {Makino}, J. 2001,
  \mnras, 321, 199

\bibitem[\protect\astroncite{{Portegies Zwart} \& {van den
  Heuvel}}{2007}]{2007Natur.450..388P}
{Portegies Zwart}, S.~F., {van den Heuvel}, E.~P.~J. 2007, \nat, 450, 388

\bibitem[\protect\astroncite{{Rycerz} et~al.}{2008a}]{2008HLA.RBS.II}
{Rycerz}, K., Bubak, M., Sloot, P. 2008a,
\newblock in Computational Science ICCS 2008 8th International Conference (eds.
  M. Bubak, G.D.v. Albada, J. Dongarra, P. Sloot), Krakow, Poland, Lecture
  Notes of Computer Science, Springer (2008), Vol. 5102, p.~217

\bibitem[\protect\astroncite{{Rycerz} et~al.}{2008b}]{2008HLA.RBS.I}
{Rycerz}, K., Bubak, M., Sloot, P. 2008b,
\newblock in Parallel Processing and Applied Mathematics 7th International
  Conference, PPAM 2007, (eds. R. Wyrzykowski, J. Dongarra, K. Karczewski, J.
  Wasniewski), Gdansk, Poland, Lecture Notes of Computer Science, Springer
  (2008), Vol. 4957, p.~780

\bibitem[\protect\astroncite{{Salpeter}}{1955}]{1955ApJ...121..161S}
{Salpeter}, E.~E. 1955, \apj, 121, 161

\bibitem[\protect\astroncite{{Sills} et~al.}{2003}]{2003NewA....8..605S}
{Sills}, A., {Deiters}, S., {Eggleton}, P., {Freitag}, M., {Giersz}, M.,
  {Heggie}, D., {Hurley}, J., {Hut}, P., {Ivanova}, N., {Klessen}, R.~S.,
  {Kroupa}, P., {Lombardi}, Jr., J.~C., {McMillan}, S., {Portegies Zwart}, S.,
  {Zinnecker}, H. 2003, New Astronomy, 8, 605

\bibitem[\protect\astroncite{{Springel} et~al.}{2001}]{2001NewA....6...79S}
{Springel}, V., {Yoshida}, N., {White}, S.~D.~M. 2001, New Astronomy, 6, 79

\bibitem[\protect\astroncite{{Suzuki} et~al.}{2007}]{2007ApJ...668..435S}
{Suzuki}, T.~K., {Nakasato}, N., {Baumgardt}, H., {Ibukiyama}, A., {Makino},
  J., {Ebisuzaki}, T. 2007, \apj, 668, 435

\bibitem[\protect\astroncite{{Williams} et~al.}{2000}]{2000prpl.conf...97W}
{Williams}, J.~P., {Blitz}, L., {McKee}, C.~F. 2000, Protostars and Planets IV,
   97

\bibitem[\protect\astroncite{{Wisdom} \& {Holman}}{1991}]{1991AJ....102.1528W}
{Wisdom}, J., {Holman}, M. 1991, \aj, 102, 1528

\end{thebibliography}
\end{document}